\documentclass[12pt]{article}
\usepackage{setspace}
\usepackage{graphicx}
\usepackage{amsfonts}
\usepackage{amsmath}
\usepackage{mathrsfs}
\def\beq{\begin{equation}}
\def\eeq{\end{equation}}
\def\bea{\begin{eqnarray}}
\def\eea{\end{eqnarray}}
\def\nn{\nonumber}
\def\ba{\begin{array}}
\def\ea{\end{array}}

\def\de{\delta}
\def\one{1\hskip -1mm{\rm l}}

\begin{document}

\begin{center}
{\large \bf \sf
Spectral properties of supersymmetric Polychronakos spin chain \\
associated with $A_{N-1}$ root system}

\vspace{1.3cm}

{\sf B. Basu-Mallick\footnote{ 
e-mail address: bireswar.basumallick@saha.ac.in}
and Nilanjan Bondyopadhaya\footnote{e-mail address:
nilanjan.bondyopadhaya@saha.ac.in } }

\bigskip

{\em Theory Group, \\
Saha Institute of Nuclear Physics, \\
1/AF Bidhan Nagar, Kolkata 700 064, India } \\
\bigskip

\end{center}

\noindent {\bf Abstract}

\noindent
By using the exact partition function of $su(m|n)$ Polychronakos 
spin chain associated with $A_{N-1}$ root system, we study some 
statistical properties of the related spectrum. 
It is found that the corresponding energy level density satisfies 
the Gaussian distribution and the cumulative distribution of spacing 
between consecutive energy levels obeys a certain `square root of a 
logarithm' law. 
\vskip 0.5 cm
\baselineskip 18 true pt 
\noindent {\it Keywords}: Exactly solvable quantum spin chains; 
Supersymmetry; Partition function; Level density distribution 
\newpage 

\baselineskip 24 true pt 
Exactly solvable one dimensional quantum integrable spin chains and 
dynamical models with long-range interaction [1-11] have
attracted much attention in recent years due to their appearance 
in a wide range of subjects like fractional statistics \cite{MS94,Po98}, 
quantum electric transport in mesoscopic systems [14], 
Yangian quantum group [15-17], SUSY Yang-Mills theory and 
string theory [18-20]. 
Among quantum spin systems with long-range interaction, 
the well known \mbox{spin-$\frac{1}{2}$} 
Haldane-Shastry (HS) model is introduced in an attempt to
construct an exact ground state which would coincide 
with the $U\rightarrow \infty$ limit of Gutwiller's variational 
wave function for the Hubbard model \cite{Hal88,Sh88}. 
A natural $su(m)$ generalization of
this exactly solvable HS model is constructed
by using the `spin' exchange operator associated with
the fundamental representation of $su(m)$ algebra \cite{Ka92,Ha92}.
Subsequently, it is realized that such HS spin chain may be reproduced 
from the trigonometric spin Calogero-Sutherland model
by applying the `freezing trick', which basically uses the fact that
spin and dynamical degrees of freedom of the latter model decouple 
from each other for large values of the coupling constant. 
Furthermore, a new quantum spin chain with long-range 
interaction is obtained by applying this freezing trick to 
the case of rational spin Calogero model \cite{Po93}. 
Lattice sites of this spin chain, 
which are inhomogeneously distributed on a line, are determined 
through the zeros of the Hermite polynomial \cite{Fr93}. 
This quantum integrable as well as exactly solvable spin system  
is usually known as Polychronakos 
or Polychronakos-Frahm spin chain in the literature.  
Both HS and Polychronakos spin chains admit $su(m|n)$ supersymmetric
extensions, in which each site is occupied by either one of the $m$
type of bosonic states or one of the $n$ type of fermionic states [21-27]. 
Such supersymmetric spin chains play a role
in describing some strongly correlated systems in condensed matter physics,
where holes moving in the dynamical back ground of spins behave as bosons,
and spin-1/2 electrons behave as fermions \cite{Sc97, Ar0106}.

It should be noted that, the above mentioned 
HS and Polychronakos spin chains  (along with their supersymmetric extensions)
are all related to the $A_{N-1}$ type of root system, for which 
the interaction between any two 
spins depends only on the difference of their site coordinates. 
Variants of these spin chains associated with
other root systems,  endowed with 
more general type of interactions, have 
also been studied in the literature [30-36]. 
   
By applying the method of freezing trick, it has become possible to 
compute the exact partition functions of $su(m)$ Polychronakos 
and $su(m)$ HS spin chains associated with  $A_{N-1}$ root system 
\cite{Po94,FG05}, 
$su(m|n)$ supersymmetric extensions of these spin chains 
\cite{BMUW99,BMN06}, 
and variants of such   
spin chains related to other root systems [33-36].
These exact partition functions have turned out to be a very 
efficient tool for studying some statistical
properties of the related energy spectra 
like level density distribution and distribution of
spacing between consecutive energy levels.  It is found that,
for sufficiently large number of lattice sites, 
 the energy level density of this type of 
spin chain follows the Gaussian distribution with high
degree of accuracy [33-38,25]. It is also observed that, distribution of
spacing between consecutive energy levels for such 
integrable spin chains is not of Poisson type, as may be expected
due to a well-known conjecture of Berry and Tabor \cite{BT77}.
However it appears that, even though the exact partition function 
of $su(m|n)$ supersymmetric Polychronakos spin chain associated with 
$A_{N-1}$ root system has been derived by using the freezing trick
\cite{BMUW99}, statistical properties of the related spectrum 
like distributions of energy level density and 
spacing between consecutive energy levels have not been analyzed till now. 
The purpose of this letter is to study these spectral properties of the  
supersymmetric Polychronakos spin chain by using its exact partition 
function.

The Hamiltonian of the $su(m|n)$ supersymmetric Polychronakos spin chain 
associated with the $A_{N-1}$ root system is given by \cite{BMUW99}
\beq
 \mathcal{H}^{(m|n)} ~=~  \sum_{ 1 \leq j <k \leq N } \,
\frac{ \left( 1- {\hat P}^{(m|n)}_{jk} \right)}{ \left ( x_j 
- x_k \right)^2 }
\, ,
\label{a1}
\eeq
where $x_j$'s are the zeros of the  $N$-th order Hermite polynomial,
and the supersymmetric exchange operator 
${\hat P}^{(m|n)}_{jk}$ is defined as 
\beq
 {\hat P}^{(m|n)}_{jk}
  \equiv
  \sum_{\alpha , \beta = 1}^{m+n}
  C_{j, \alpha}^\dagger \, C_{k, \beta}^\dagger  \,
  C_{j , \beta} \, C_{k, \alpha} .
\label{a3}
\eeq
Here creation-annihilation operators like $C_{j,\alpha}^\dagger$ and 
$C_{j, \alpha}$ are bosonic for $\alpha \in \{1,2, \dots, m \}$
and fermionic for $\alpha \in \{ m+1,m+2, \dots, m+n \}$, 
and these operators act on a restricted Hilbert space  
satisfying the constraint:
\begin{equation*}
  \sum_{\alpha=1}^{m+n} C_{j, \alpha}^\dagger \, C_{j, \alpha} = 1 ,
\end{equation*}
for all $j$. It may be noted that for the special case like $m \neq 0,n=0$
or $m=0, ~n \neq 0$,
i.e. when all degrees of freedom are either bosonic or fermionic,
$\mathcal{H}^{(m|n)}$ in Eq.~(\ref{a1}) reduces to the
non-supersymmetric Polychronakos spin chain, for which statistical
properties of the spectrum have been studied very recently \cite{BFGRepl}.
In this article we shall analyze the spectrum of 
$\mathcal{H}^{(m|n)}$ in the purely supersymmetric case,
when both $m$ and $n$ take nonzero values.

As shown in Ref.~\cite{Ba99}, 
the supersymmetric exchange operator (\ref{a3})
can be mapped to an `anyon like' representation of the permutation algebra 
on an appropriate spin space. 
By using this mapping and subsequently applying the freezing trick, 
the exact partition function
 of $su(m|n)$ Polychronakos spin chain has been derived as \cite{BMUW99} 
 \beq
  Z^{(m|n )}_N (q) \, ~=~ \,  \sum_{ \sum_{i=1}^m a_i + 
  \sum_{j=1}^n b_j =N } ~
 \frac{  { (q)_N   \, \cdot \, q^{  \sum_{j=1}^n  \frac{b_j (b_j -1 ) }{ 2}  } }
 }{ (q)_{a_1} (q)_{a_2} \cdots  (q)_{a_m} \cdot (q)_{b_1} (q)_{b_2}
  \cdots  (q)_{b_n}  } \, ,
\label {a4}
\eeq
where $a_i$'s and $b_i$'s are non-negative integers, 
$ q\equiv e^{- 1/kT}$ and  
the notation: $(q)_N \equiv (1-q) (1-q^2) \cdots (1-q^N) \, $
(with $(q)_0 \equiv 1 $) is used. 
Exploiting a connection between this  partition function 
and supersymmetric Schur polynomials, it is found that
all energy eigenvalues of $su(m|n)$ Polychronakos spin chain
can be expressed in terms of 
`motifs', which are represented by arrays of binary digits 
like `$0$' and `$1$' \cite{HBM00}. 
For a spin chain with $N$ number of lattice sites, each motif is 
represented by an array of $N-1$ number of binary digits.  
Thus we may write such a motif as  
$(\delta_1,\delta_2,\dots,\delta_{N-1})$, 
where $\delta_i \in [0,1]$. 
These motifs  characterize a class of
irreducible representations of  $Y(gl(m|n))$ Yangain quantum group, 
which span the Fock space of $su(m|n)$ supersymmetric Polychronakos spin chain.
Since $Y(gl(m|n))$ Yangain is the symmetry algebra of this 
spin chain, all of its eigenstates associated with any particular 
irreducible representation or corresponding motif 
yield the same energy eigenvalue. 
One can express the energy eigenvalue of such 
degenerate eigenfunctions associated with the motif 
$\delta \equiv (\delta_1,\delta_2,\dots,\delta_{N-1})$ as 
\beq
 E (\de)  ~=~ \sum_{j=1}^{N-1} ~j \, \de_j \, . \label{b22} 
\eeq
It should be noted that, in contrast to the non-supersymmetric case, 
there is no restriction in arranging $1$'s 
or $0$'s in a motif associated with the supersymmetric 
Polychronakos spin chain \cite{HBM00}. 
Consequently, some additional motifs and energy levels appear 
in the spectrum of this supersymmetric spin chain in comparison with its
non-supersymmetric counterpart. In fact,   
for this supersymmetric spin chain with $N$ number of 
lattice sites, one can construct $2^{N-1}$ number 
of motifs by filling up $N-1$ positions arbitrarily with either $0$ or $1$. 
By using Eq.~(\ref{b22}),  
it is easy to check that the motif 
$(0,0, \dots, 0)$ yields the ground state with zero energy 
and the motif $(1,1,\dots,1)$ yields the highest excited 
state with energy $N(N-1)/2$. Moreover, depending on the choice 
of $\delta$, $E (\de)$ in Eq.~(\ref{b22}) 
can take any integer value within the range $0$ to $N(N-1)/2$.
Consequently, the spectrum of $su(m|n)$ supersymmetric Polychronakos 
spin chain would be equispaced in nature.

Since $Z^{(m|n )}_N (q)$ in Eq.~(\ref {a4}) 
represents the partition function of a finite spin system 
with integer eigenvalues, 
it should be expressed as a polynomial function of $q$. Indeed, 
with the help of a symbolic software package like Mathematica, 
we can explicitly write down $Z^{(m|n )}_N (q)$ 
as a polynomial of $q$ for wide range of 
values of $m$, $n$ and $N$. 
If the term $q^{E_i}$ appears in such a polynomial,
then $E_i$ will represent an energy
eigenvalue and the coefficient of $q^{E_i}$ will determine the 
degeneracy factor corresponding to this energy level. Let us denote 
this degeneracy factor or `level density' associated with the  
energy level $E_i$ as $D^{(m|n)}(E_i)$.  
Since the sum of these level densities for
the full spectrum is not normalized to unity, 
we define normalized level density as
\beq
\mathcal{D}^{(m|n)}(E_i)=\frac{1}{(m+n){}^N}\,D^{(m|n)}(E_i),
\label{a5}
\eeq
such that $\sum_{E_i} \mathcal{D}^{(m|n)}(E_i)= 1$.
In the following, we shall use this $\mathcal {D}^{(m|n)}(E_i)$  
to study some spectral properties of 
the supersymmetric Polychronakos spin chain. 

Let us investigate at first whether, in analogy with the case 
of many other quantum integrable spin chains with long-range interaction, 
the energy level density of $su(m|n)$ supersymmetric Polychronakos spin chain 
can be described by the Gaussian distribution. 
The form of normalized Gaussian 
distribution is given by 
\beq 
G(E)=\frac{1}{ \sqrt{2\pi} \sigma}
\exp\left[- \frac{{(E-\mu )}^2}{2 \sigma^2}\right],
\label{a6}
\eeq
where $\mu $ and $\sigma $ denote the mean 
value and standard deviation respectively.  For the case of 
supersymmetric Polychronakos spin chain,
these parameters are evidently related to the Hamiltonian $ {\cal H}^{(m|n)}$
in Eq.~(\ref{a1}) as 
\bea
~~~~~~~~~~~~~~~~~~~~\mu=\frac{tr\left[ 
{\cal H}^{(m|n)}\right]}{{(m+n)}^N}\,, ~~~~~\sigma^2=
  \frac{tr\left[({\cal H}^{(m|n)})^2 
\right]}{{(m+n)}^N} \, - \, \mu^2. 
\label{7}
\eea
In order to compare $\mathcal{D}^{(m|n)}(E_i)$ with $G(E_i)$, 
we have to compute the parameters $\mu$ and $\sigma$ 
in terms of $m$, $n$ and $N$.
For this purpose, let us consider 
some trace formulas involving supersymmetric 
exchange operators like \cite{BMN06}
\bea
\begin{aligned}
~~~& tr\left[ 
(\hat{P}_{ij}^{(m|n)})^2\right]=tr\left[ \one \right]=
s^N,~~~~tr\left[ \hat{P}_{ij}^{(m|n)} 
\right]=s^{N-2}t \, ,
\nn \\   
~~~&tr \left[ \hat{P}_{ij}^{(m|n)}\hat{P}_{il}^{(m|n)} \right] 
=tr\left[ \hat{P}_{ij}^{(m|n)}
\hat{P}_{jl}^{(m|n)}\right] =tr\left[ 
\hat{P}_{ij}^{(m|n)}\hat{P}_{kj}^{(m|n)}\right]=s^{N-2},
\nn \\
~~~&tr\left[\hat{P}_{ij}^{(m|n)}\hat{P}_{kl}^{(m|n)}\right]=s^{N-4}t^2 ,
\nn
\end{aligned}
\eea
\addtocounter{equation}{2}
\noindent where $s=m+n$, $t=m-n$ and $i,j,k,l$ are all different indices. 
With the help of these 
trace formulas, $\mu$ and $\sigma$ in 
Eq.~(\ref{7}) can be expressed as 
\bea
\mu &=& \frac{1}{2s^2}(s^2-t) \sum_{\substack{i,j=1 \\ (i \neq j)}}^N
 \frac{1}{x_{ij}^2}\, , \nn \\
 \sigma^2 &=& \frac{s^4-t^2}{2s^4}\sum_{\substack{i,j=1 \\ (i \neq j)}}^N
 \frac{1}{x_{ij}^4}+\frac{s^2-t^2}{s^4}\sum^N_{\substack{i,j,k =1 \\ 
(i \neq j \neq k \neq i )}} \frac{1}{x_{ij}^2x_{jk}^2}\, ,
\label{a8++}
\eea
where $x_{ij} \equiv x_i -x_j$.
These two expressions encompass different types of summations
involving the zeros of the $N$-th order Hermite polynomial.
It is well known that, the zeros of the $N$-th order Hermite polynomial
satisfy an identity of the form \cite{A78}
\beq
\sum_{\substack{k=1\\ (k \neq i)}}^N\frac{1}{x_{ik}}=x_i \, .
\label{a9}
\eeq
Starting from this simple identity, it is possible  
to derive many other more complicated form of identities 
involving the zeros of the Hermite polynomial. 
In particular, one can show that \cite{A78}
\bea
~~~~~~~~~~~~~
\sum_{\substack{i,j=1 \\ (i \neq j)}}^N \frac{1}{x_{ij}^2} =
\frac{1}{2}N(N-1) \, , ~~~~
\sum_{\substack{i,j=1 \\ (i \neq j)}}^N \frac{1}{x_{ij}^4}= 
\frac{1}{36} N(N-1)(2N+5) \, . 
\label{a10+}
\eea
By using the identity (\ref{a9}), we also find that 
\bea
\sum^N_{\substack{i,j,k =1 \\ (i \neq j \neq k \neq i )}}
 \frac{1}{x_{ij}^2x_{jk}^2} = \frac{2}{9} N(N-1)(N-2) \, .
\label{a10}
\eea
Substituting the identities given in Eqs.~(\ref{a10+}) and 
(\ref{a10}) to 
Eq.~(\ref{a8++}),  we finally obtain 
$\mu$ and $\sigma$ in terms of $m$, $n$ and $N$ as 
\bea
 \mu &=& \frac{1}{4s^2}(s^2-t)N(N-1),  \nn \\
 \sigma^2 &=& \frac{s^4-t^2}{72s^4} N(N-1)(2N+5)+
 \frac{2(s^2-t^2)}{9s^4} N(N-1)(N-2).
\label{a16}
\eea
From the above expression it is evident that,  
$\mu$ and $\sigma$  are of the order of $N^2$ and $N^{3/2}$ respectively
at $N\rightarrow \infty $ limit. 
Interestingly, similar type of large $N$ behaviour of these parameters
have also been observed for non-supersymmetric Polychronakos 
spin chain associated with the $A_{N-1}$ root system \cite{BFGRepl} and 
non-supersymmetric as well as supersymmetric  Polychronakos
spin chains associated with the $BC_N$ root system
\cite{BFGRprb,BFGRarx}. 

Now we are in a position to estimate the agreement between
$\mathcal{D}^{(m|n)} (E_i)$ and $G(E_i)$ for the case of $su(m|n)$
supersymmetric Polychronakos spin chain. As an example,
let us consider the simplest case of $su(1|1)$ Polychronakos spin.
By using Mathematica, we can easily express 
the corresponding partition function (given in 
Eq.~(\ref{a4}) with $m=n=1$) as a polynomial of $q$
up to $N \approx 50$ and obtain 
$\mathcal{D}^{(1|1)}(E_i)$ for all values of $E_i$. 
By using such data for the specific case of $N=30$ lattice sites,
we plot $\mathcal{D}^{(1|1)}(E_i)$ versus $E_i$
 through the dotted curve in Fig.~1. Next, 
by substituting $N=30$ and $m=n=1$
in Eq.~(\ref{a16}), we get $\mu = 217.5$ and $\sigma = 46.25$. 
The continuous curve in Fig.~1 plots $G(E)$ as a function of $E$ 
for these values of $\mu$ and $\sigma$.  
Comparison between the continuous and dotted curves in 
Fig.~1 confirms that the normalized energy level density
obeys Gaussian distribution with a very high degree of accuracy.
We also calculate the mean square error (MSE) for this case
 and find it as low as  $4.74 \times 10^{-4}$. 
 Moreover, it is found that this MSE
decreases gradually with the increase of the number of lattice sites.
We have compared $\mathcal{D}^{(m|n)} (E_i)$ with $G(E_i)$ for 
many other values of $m$, $n$ and $N$, and always found excellent 
agreement between them for sufficiently large number of lattice sites. 

Next, our aim is to study the distribution of spacing between 
consecutive energy levels for the case of supersymmetric Polychronakos spin 
chain.  To this end, we define cumulative level spacing distribution
as 
$$P(s)=\int_o^s p(x)dx \, ,$$ 
where $p(x)$ is the probability density of the normalized spacing $x$ 
between consecutive (unfolded) energy
levels of the Hamiltonian. 
Unfolded energy levels are obtained by exploiting a mapping 
which  smooths out the `raw ' spectrum containing local fluctuations 
\cite{Haake}. To define such unfolding mapping, at first it is needed 
to decompose the cumulative energy level density as the sum of a 
fluctuating part and a continuous part. This 
continuous part of the cumulative energy level density
(denoted by $\eta (E)$)
is then used to transform each energy $E_i$, $i~=~1,2, \dots,l$,
into an unfolded energy $\mathcal{E}_i$ 
: $E_i \rightarrow  \mathcal{E}_i=\eta (E_i)$. 
Since we have found that, 
the energy level density of $su(m|n)$ Polychronakos spin chain
follows the continuous Gaussian distribution with very good approximation, 
$\eta(E)$ can be expressed through the error function as  
\beq
\eta (E)= \int_{-\infty}^E G(E') \, d E' = \frac{1}{2} \Bigl[ 1+ {\rm erf} \, 
\Bigl( \frac{E- \mu}{\surd2 \sigma} \Bigr) 
\Bigr].
\label{a17}
\eeq
Finally, the function $p(s_i)$
is defined as the density of normalized spacing 
$s_i=(\mathcal{E}_{i+1}-\mathcal{E}_{i})/\Delta$, where 
$\Delta=(\mathcal{E}_l-\mathcal{E}_1)/(l-1)$ is the mean
 spacing of the unfolded energy levels.  

We have already mentioned that, according to a well-known 
conjecture of Berry and Tabor,  
the density $p(s)$ of normalized  spacing 
for the case of a quantum integrable system  
should obey the Poisson's law:  $p(s)= e^{-s}$ \cite{BT77}.  
However,  some recent works reveal that a class of
quantum integrable HS and Polychronakos like spin chains exhibit 
non-Poissonian distribution of spacing between
consecutive energy levels and thus violate the Berry-Tabor conjecture 
[33-38,25].
It is also found that, for the above mentioned type of spin chains,   
 $P(s)$ obeys a certain `square root of a logarithm' law, which 
can be derived analytically 
by assuming a few simple properties of the related spectra. 
More precisely,  if any spectrum $E_{min}\equiv E_1 < \dots < E_l
 \equiv E_{max}$ obeys the following features:
\begin{description}
 \item [(i)] the energy levels are equispaced, or, equivalently 
 $E_{i+1}-E_i =\lambda$ for $1\leq i< l $,
\item [(ii)] the normalized energy level density is approximately given
by the Gaussian distribution,
\item [(iii)] two tails of the level density distribution are well spread, 
i.e, \mbox {$E_{max} - \mu , \, \mu -E_{min}  >> \sigma$}, 
\item [(iv)] $E_{min}$ and  $E_{max}$ are 
approximately symmetric with respect to the mean energy $\mu$,
namely  $| E_{max} + E_{min}-2\mu |<<E_{max}-E_{min}$, 
\end{description}
then $P(s)$ 
is approximately given by 
an analytic expression of the form \cite{BFGRprb}
\beq
{\bar P}(s) = 1- \frac{2}{\sqrt{\pi} ~s_{max}} 
\sqrt{log \Bigl( \frac{s_{max}}{s} \Bigr) } \, ,
\label{a11}
\eeq
where $s_{max}$ denotes the maximum normalized spacing, which 
can be estimated with great accuracy through the relation   
\beq
s_{max}=\frac{E_{max}-E_{min}}{\sqrt{2\pi}\sigma}.
\label{a12}
\eeq

Studying the case of $su(m|n)$ supersymmetric Polychronakos spin chain 
for a wide range of values of $m$, $n$ and $N$, 
we also find that the spacing between
consecutive energy levels does not follow the Poisson distribution. 
So it is natural to 
explore the applicability of expression (\ref{a11})  
for the present case. To this end, 
let us first check whether the energy spectrum of $su(m|n)$ 
supersymmetric Polychronakos spin chain
satisfies the four conditions 
that have been described in the previous paragraph.
We have already found  that this spectrum 
is equally spaced with unit interval 
and the level density follows Gaussian distribution with 
good approximation for sufficiently large values of $N$.
So this spectrum evidently satisfies the first two conditions. 
From Eq.~(\ref{a16}) and the expressions of lowest and highest 
energy levels given by $E_{min}=0$ and $E_{max}= N(N-1)/2$ respectively, 
it follows that both  
$(\mu -{E}_{min} )/\sigma$ and  $({E}_{max}-\mu )/\sigma$
vary as $\sqrt{N}$ when $N\rightarrow \infty$.
Therefore, this spectrum conforms to the third condition. At 
$N\rightarrow \infty$ limit, it is also easy to find that 
\beq
| {E}_{min}+{E}_{max}-2\mu | =X(m,n)({E}_{max}
- {E}_{min}),
\label{a13}
\eeq
where $X(m,n)=\frac{|m-n|}{(m+n)^2}$.
Note that $X(m,n)$ becomes zero for $m=n$
and  takes a finite nonzero value for $ m \neq n$.
Consequently,  the fourth condition is satisfied only for the case $m=n$. 
However it can be shown that, if one drops this forth condition, 
Eq.~(\ref{a11}) still holds within a slightly 
smaller range of $s$ \cite{BFGRarx}. Therefore, it is natural to expect that 
$P(s)$ would follow the analytical expression ${\bar P}(s)$ (\ref{a11}) 
in the case $su(m|n)$ Polychronakos spin chain 
for all possible values of $m, ~n$ and sufficiently large values of $N$. 

To support the above conclusion, we study numerically the nature of $P(s)$
with different values of $m$, $n$ and $N$.
For example, we may consider the particular case of 
$su(1|1)$ spin chain with $N=30$ lattice sites. 
For this case, $P(s)$ and ${\bar P}(s)$ are drawn  
as the  dotted line and the continuous line respectively in Fig.~2. 
From this figure it is evident that,   
  the cumulative distribution of spacing matches with 
     ${\bar P}(s)$  extremely well.  The MSE for this case 
is obtained as $2.628 \times 10^{-6}$. Next, we consider the case of
$su(2|1)$ spin chain with $N=30$ lattice sites. The corresponding  $P(s)$
  and ${\bar P}(s)$  are plotted as the dotted line and the continuous line 
  respectively in Fig.~3.
Again, a very good agreement is found between these two lines with 
MSE given by $4.724 \times 10^{-4}$. By studying such particular cases
we conclude that, similar to the case of other quantum integrable
spin chains with long-range interaction that have been studied so far, 
the cumulative distribution of spacing between consecutive 
energy levels of the supersymmetric Polychronakos spin chain 
follows the expression (\ref{a11}) with remarkable accuracy.


\newpage
\begin{figure}[h]
\centering 
\hskip -.6 cm 
\includegraphics [scale=0.60,angle=270]{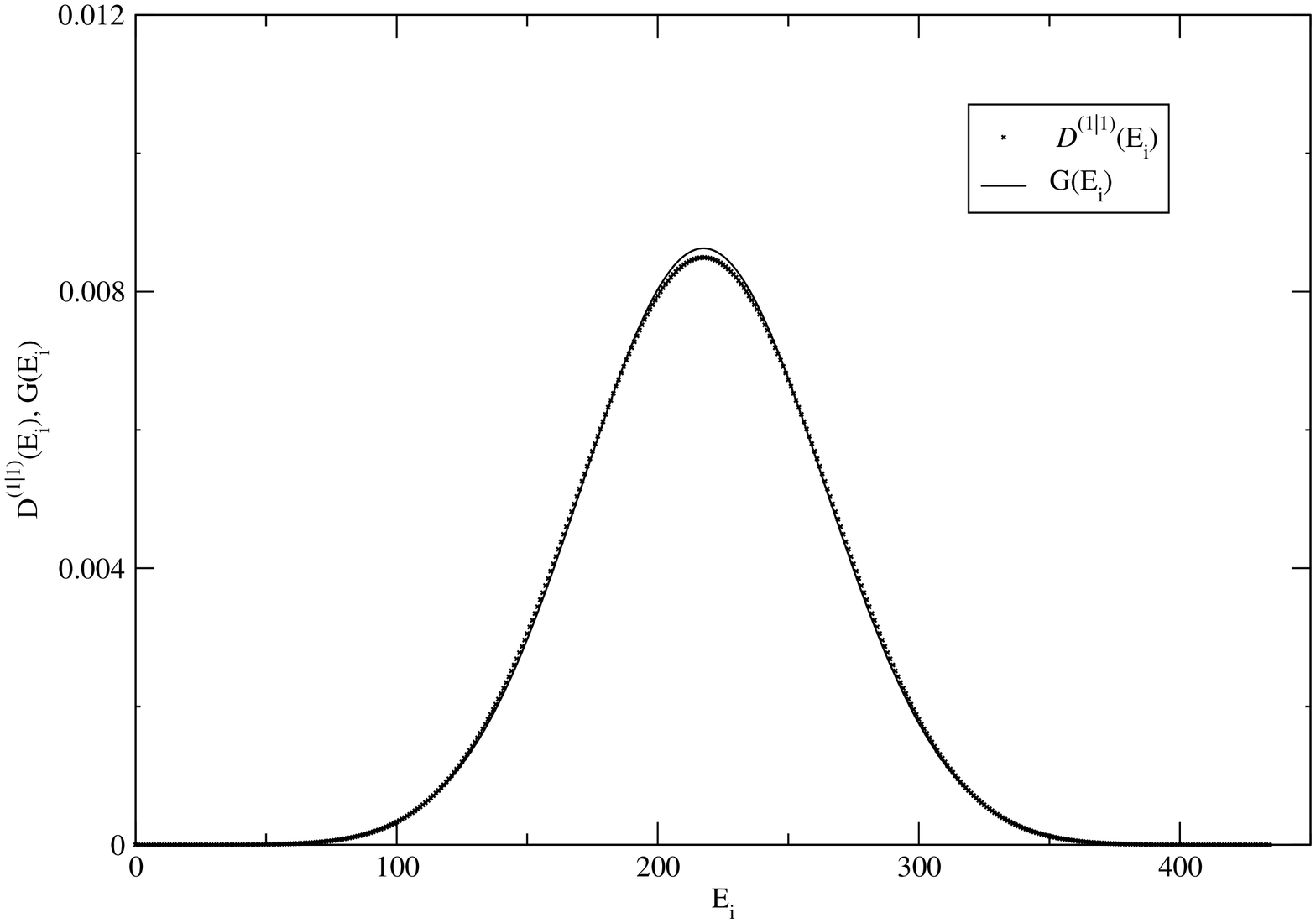}
\caption{Plot of the degeneracies $\mathcal{D}^{(1|1)}(E_i)$ versus 
 energies $E_i$ (dotted line) for the case of 
$su(1|1)$ Polychronakos spin chain with $N=30$, 
and its comparison with the Gaussian distribution (continuous line). }

\end{figure}
\newpage
\begin{figure}[h]
\centering 
\hskip -.25 cm 
\includegraphics[scale=0.60,angle=270]{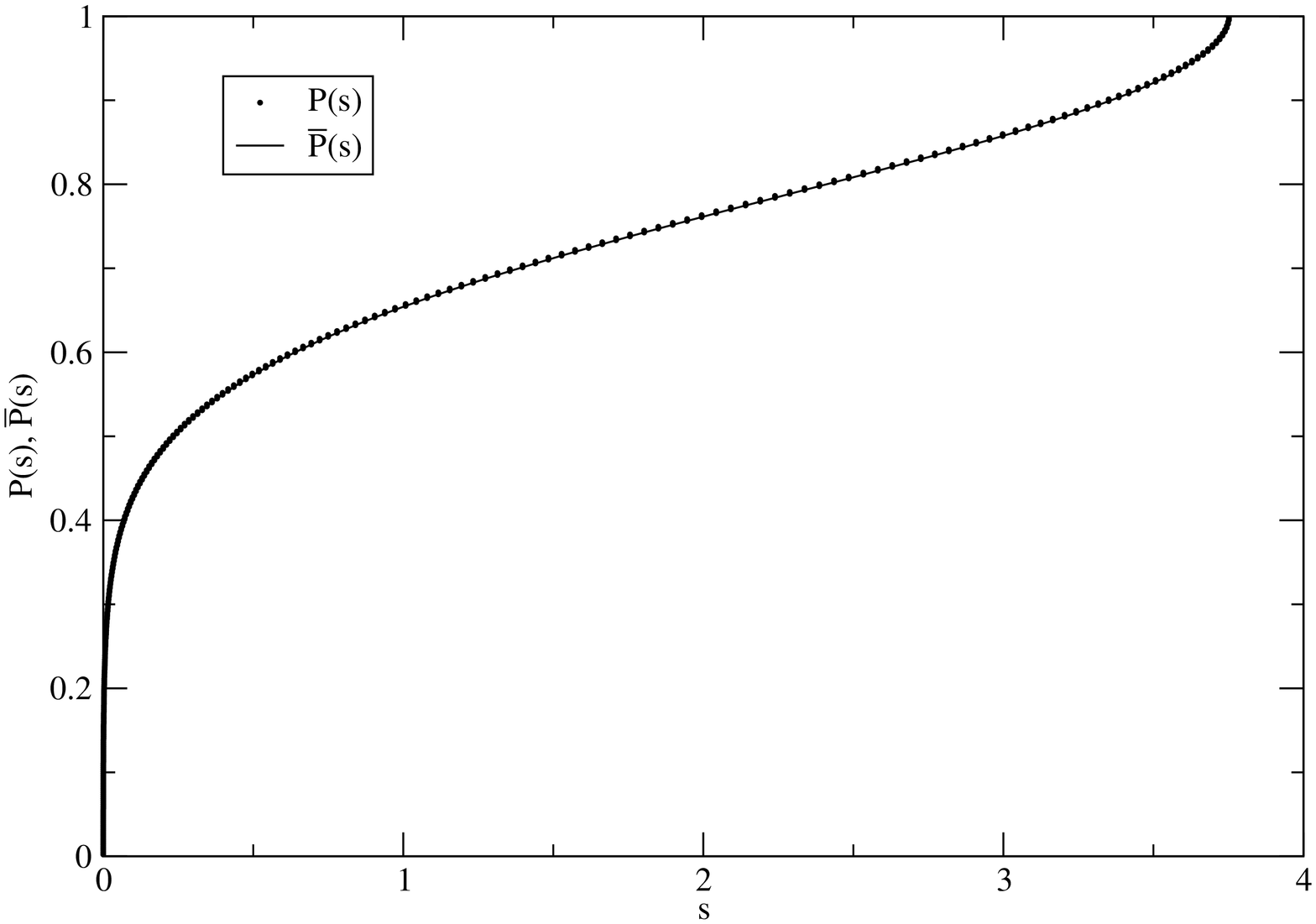}
\caption{ The dotted curve represents  the cumulative distribution 
of spacing between consecutive energy levels 
for $su(1|1)$ spin chain with $N=30$,  and the continuous 
curve represents the corresponding  ${\bar P}(s)$. }
\end{figure}

\newpage
\begin{figure}[h]
\centering 
\hskip -.25 cm 
\includegraphics[scale=0.60,angle=270]{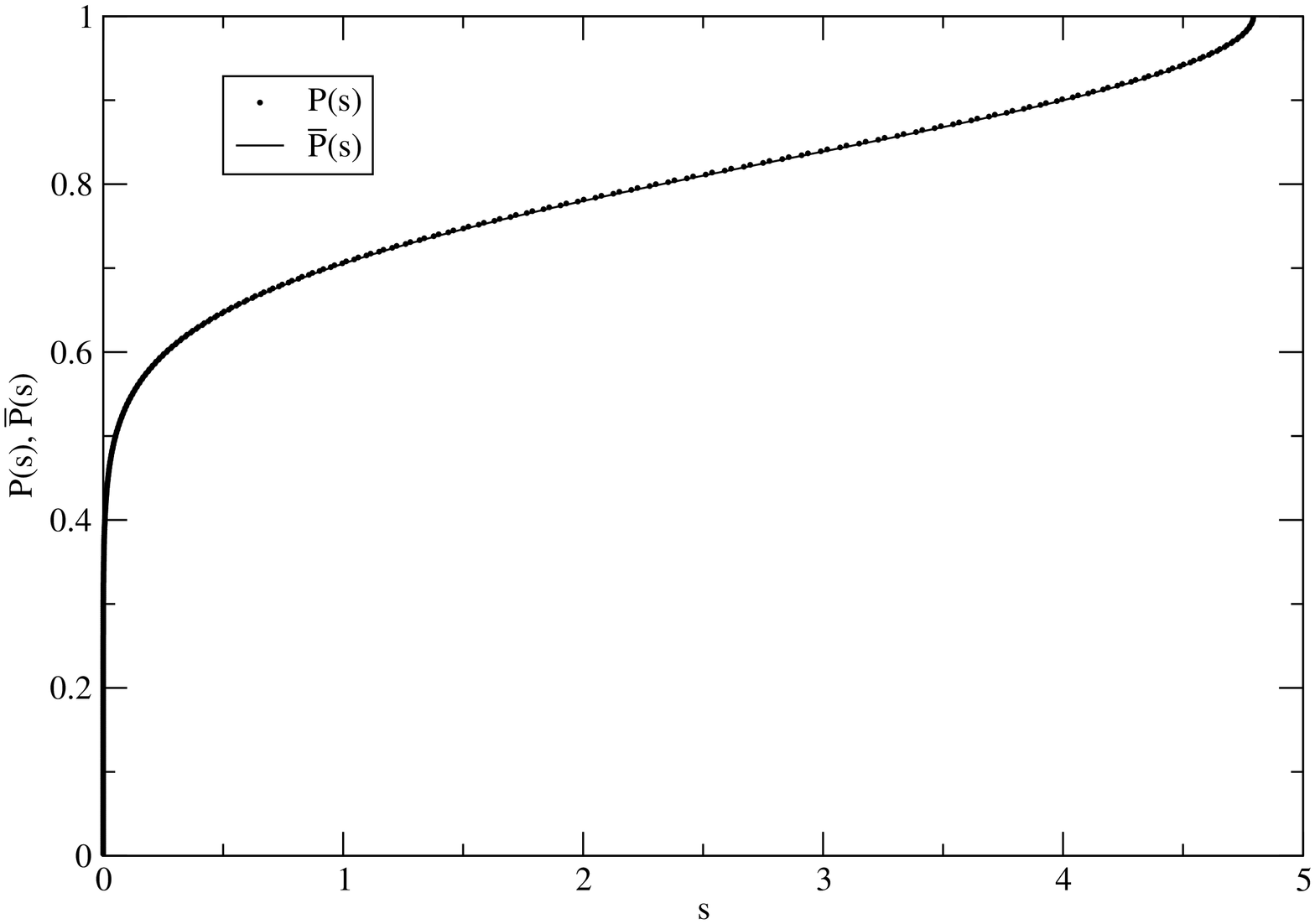}
\caption{ The dotted curve represents  the cumulative distribution
of spacing between consecutive energy levels
for $su(2|1)$ spin chain with $N=30$,  and the continuous
curve represents the corresponding  ${\bar P}(s)$. }

\end{figure}

\end{document}